\documentclass[12pt,english,english,english,english,english,english]{article}
\usepackage[T1]{fontenc}
\usepackage[latin1]{inputenc}
\usepackage{babel}

\makeatletter

\providecommand{\LyX}{L\kern-.1667em\lower.25em\hbox{Y}\kern-.125emX\@}

\usepackage[T1]{fontenc}
\usepackage[latin1]{inputenc}
\usepackage{babel}

\makeatletter

\usepackage[T1]{fontenc}
\usepackage[latin1]{inputenc}
\usepackage{babel}

\makeatletter

\usepackage[T1]{fontenc}
\usepackage[latin1]{inputenc}
\usepackage{babel}

\makeatletter

\usepackage[T1]{fontenc}
\usepackage[latin1]{inputenc}
\usepackage{babel}

\makeatletter

\usepackage[T1]{fontenc}
\usepackage[latin1]{inputenc}
\usepackage{babel}



\makeatother

\makeatother

\makeatother

\makeatother

\makeatother
\begin{document}

\title{The thermal energy of a scalar field in a one-dimensional compact
space}

\author{E. Elizalde\( ^{\dagger } \)\\
 Institut d'Estudis Espacials de Catalunya (IEEC/CSIC) \\
 Edifici Nexus, Gran Capit\`{a} 2-4, 08034 Barcelona, Spain; \\
 Departament d'Estructura i Constituents de la Mat\`{e}ria \\
 Facultat de F\'{\i}sica, Universitat de Barcelona \\
 Av. Diagonal 647, 08028 Barcelona, Spain\\
 and \\
 A. C. Tort\( ^{\ddagger ,\star } \)\\
 Departamento de F\'{\i}sica Te\'{o}rica - Instituto de F\'{\i}sica
\\
 Universidade Federal do Rio de Janeiro\\
 C.P. 68528, 21945-970 Rio de Janeiro, Brazil}

\date{\today{}}

\maketitle
\begin{abstract}
\noindent We discuss some controverted aspects of the evaluation of
the thermal energy of a scalar field in a one-dimensional compact
space. The calculations are carried out using a generalised zeta function
approach. \vskip 2cm
\end{abstract}
\( ^{\dagger } \) e-mail: elizalde@ieec.fcr.es \\
 \( ^{\ddagger } \) e-mail: tort@if.ufrj.br \\
 \( ^{\star } \) Present address: Institut d'Estudis Espacials de
Catalunya (IEEC/CSIC) Edifici Nexus, Gran Capit\`{a} 2-4, 08034 Barcelona,
Spain. E-mail address: visit11@ieec.fcr.es

\section{Introduction}

In a recent paper Brevik, Milton and Odintsov \cite{Brevik al 02}
calculated the thermal energy associated with several types of quantum
fields living in \( S^{1}\times S^{d-1} \) geometries with \( d=2,3 \).
In particular the thermal energy of a quantum neutral scalar field
defined on a one-dimensional compact space at finite temperature (\( S^{1}\times S^{1} \))
was found by those authors to depend quadratically and also linearly
on the temperature in the high-temperature limit. Their result reads\begin{equation}
\label{BMO result}
E\left( \beta \right) \simeq \frac{1}{24a}\left( \frac{2\pi a}{\beta }\right) ^{2}-\frac{1}{2\beta },\, \, \, \, \, \, \, \, \, \, \, \, \, \, \, a/\beta \, \gg \, 1,
\end{equation}
 where \( \beta  \) is the reciprocal of the thermal equilibrium
temperature \( T \). This expression is in agreement with the corresponding
one obtained by Kutasov and Larsen \cite{Kutasov and Larsen 2000}
(in the high-temperature limit) but at variance with the one obtained
by Klemm \emph{et al} \cite{Klemm at al 2000}. These last authors
justify their result by arguing that the contribution of the zero
modes becomes significant only when the saddle-point approximation
does not hold anymore. This disagreement has arisen some doubts concerning
the correctness of Brevik \emph{et al}'s evaluation. Some other authors
also strongly support the necessity to get rid of the zero mode (see,
for instance, \cite{ovm}), a common practice also in e.g. zeta function
regularization. However, it has been argued by Dowker that the correct
result for the thermal energy for this particular arrangement of quantum
field and geometry contains a positive, linearly-dependent term in
the temperature that in the high-temperature limit cancels exactly
the corresponding negative term in (\ref{BMO result}), see \cite{Dowker 2002}
and references therein. This same result can be inferred from Cardy's
analysis of partition functions for conformal field theories in higher
dimensions \cite{Cardy1991}. Besides a fundamental problem concerning
the associated entropy function, this disagreement may have consequences
for the Verlinde-Cardy relations \cite{Verlinde,Cardy86}.

The purpose of this brief note is to examine such questions a little
bit further. We will see that a generalised zeta function approach
to the problem shows that: (i) there is indeed a contribution to the
thermal energy of the spatial zero mode as calculated independently
by Dowker \cite{Dowker84} and Dowker and Kirsten \cite{DowkerK99},
and also Cardy \cite{Cardy1991}; (ii) though the thermal energy is
independent of the scaling mass inherent to zeta function regularization,
the free energy and the entropy associated with the spatial zero mode
depend on the scaling mass; and finally, (iii) that there is no sense
in discarding the terms that depend on the scaling mass, since this
would lead to a violation of the fundamental thermodynamical relations;
in fact, these need the inclusion of a dimensionful parameter to go
with \( \beta  \) and provide the unit scale in such a way that the
laws of thermodynamics be obeyed by the thermodynamics of the zero
mode. Or, to put it in other words, it does not seem to have a sense
to choose a particular, convenient value of the scaling mass at this
stage of the procedure, any such value being inescapably arbitrary.
The full resolution of the problem can only be obtained after carrying
out a compulsory renormalization procedure which makes contact with
physical experiment.

\section{Zeta function approach to the partition function}

We begin by briefly sketching a generalised zeta function \cite{Elisalde94}
approach to the partition function of a quantum scalar field in thermal
equilibrium with a heat bath at a fixed temperature \( \beta ^{-1} \).

The zeta function for conformally coupled massless scalars fields
living on a compact \( d_{R} \)-dimensional manifold at finite temperature
reads\begin{eqnarray}
\zeta \, \left( s;\frac{\partial ^{2}+\xi R}{\mu ^{2}}\right)  & = & \mbox {Tr}\left( \frac{\partial ^{2}+\xi R}{\mu ^{2}}\right) ^{-s}\nonumber \\
 & = & \mu ^{2s}\sum _{\lambda }\left( \omega ^{2}_{\lambda }\right) ^{-s},\label{thermal zeta}
\end{eqnarray}
 where \( \partial ^{2}:=\nabla ^{2}+\partial ^{2}/\partial ^{2}\tau  \),
\( \mu  \) is a scaling mass, \( \xi =\left( D-2\right) /4\left( D-1\right)  \)
is the conformal parameter and \( R \) is the Ricci curvature scalar.
Here \( D=d_{R}+1 \) where \( d_{R} \) denotes the number of dimensions
of the manifold. The eigenvalues \( \omega ^{2}_{\lambda } \) of
the Euclidean d'Alembertian operator are given by\begin{equation}
\label{dAlembert eigenvalues}
\omega ^{2}_{\lambda }=\left( \frac{2\pi n}{\beta }\right) ^{2}+\left( \frac{M^{2}_{\ell }}{a^{2}}\right) ^{2},
\end{equation}
 where, in principle, \( n=0,\pm 1,\pm 2,\ldots  \). Here \( M^{2}_{\ell }/a^{2} \)
denote the eigenvalues of the Laplace operator on the \( d_{R} \)-dimensional
manifold. For a bosonic neutral scalar field the partition function
can be obtained from the zeta function through \begin{equation}
\log \, Z\left( \beta \right) =\frac{1}{2}\zeta ^{\prime }\left( s=0;\frac{\partial ^{2}+\xi R}{\mu ^{2}}\right) .
\end{equation}

We can separate the infinite temperature sector, which corresponds
to \( n=0 \), i.e., the zero mode associated with the compactified
Euclidean temporal dimension, from the finite temperature sector and
at the same time isolate the spatial zero mode in the following way.
Suppose that the spatial zero mode corresponds to \( \ell =0 \) and
the remaining \( \ell  \) are shifted to \( \ell +1 \) such that
\( \ell =0,1,2,\ldots  \) with an appropriate degeneracy factor taken
into account. For convenience, however, suppose initially that the
spatial zero mode is absent. Then taking Eq. (\ref{dAlembert eigenvalues})
into (\ref{thermal zeta}), performing a Mellin transform, and making
use of the Poisson sum formula (or Jacobi theta function identity)
\begin{equation}
\label{po1}
\sum _{n=-\infty }^{\infty }e^{-n^{2}\pi \tau }=\frac{1}{\sqrt{\tau }}\sum _{n=-\infty }^{\infty }e^{-n^{2}\pi /\tau },
\end{equation}
 with \( \tau =4\pi t/\beta ^{2} \), we have \begin{eqnarray}
\zeta \, \left( s;\frac{\partial ^{2}+\xi R}{\mu ^{2}}\right)  & = & \frac{\mu ^{2s}}{\Gamma \left( s\right) }\int _{0}^{\infty }dt\, t^{s-1}\sum _{\ell =0}^{\infty }D_{\ell }e^{-\left( M^{2}_{\ell }/a^{2}\right) t}\sum _{n=-\infty }^{\infty }e^{-\left( 4\pi ^{2}n^{2}/\beta ^{2}\right) t}\nonumber \\
 & = & \frac{\mu ^{2s}}{\Gamma \left( s\right) }\frac{\beta }{\sqrt{4\pi }}\sum _{\ell =0}^{\infty }D_{\ell }\int _{0}^{\infty }dt\, t^{s-3/2}e^{-\left( M^{2}_{\ell }/a^{2}\right) t}+\frac{\mu ^{2s}}{\Gamma \left( s\right) }\frac{\beta }{\sqrt{\pi }}\sum _{\ell =0}^{\infty }D_{\ell }\nonumber \\
 & \times  & \sum _{n=1}^{\infty }\int _{0}^{\infty }dt\, t^{s-3/2}e^{-n^{2}\beta ^{2}/4t-M^{2}_{\ell }/a^{2}t},
\end{eqnarray}
 where \( D_{\ell } \) is the degeneracy factor. It is not hard to
see that the spatial zero mode can be taken into account by considering
the additional term\begin{equation}
\zeta _{Z.M.}\, \left( s;\frac{\partial ^{2}+\xi R}{\mu ^{2}}\right) =\frac{\mu ^{2s}}{\Gamma \left( s\right) }\frac{\beta }{\sqrt{\pi }}\sum _{n=1}^{\infty }\int _{0}^{\infty }dt\, t^{s-3/2}e^{-n^{2}\beta ^{2}/4t}.
\end{equation}
 The contribution of the spatial zero mode as it stands is divergent,
the origin of this divergence being the simultaneous consideration
of two zero modes, the temporal and the spatial ones. Nevertheless,
we can regularise it and extract its finite contribution in several
ways. One of this ways is to apply Eq. (\ref{po1}) and discard terms
that do not depend on \( \beta  \) or contribute to the partition
function with a term which is linear in \( \beta  \). We obtain\begin{equation}
\zeta _{Z.M.}\, \left( s;\frac{\partial ^{2}+\xi R}{\mu ^{2}}\right) =2\left( \frac{\mu \beta }{2\pi }\right) ^{2s}\zeta _{R}\left( 2s\right) .
\end{equation}
 Since \( \zeta _{R}\left( 0\right) \neq 0 \), we can expect a scaling
mass dependence of some or all thermodynamical quantities associated
with the spatial zero mode. The derivative at \( s=0 \) reads\begin{equation}
\zeta ^{\, \prime }_{Z.M.}\, \left( s=0;\frac{\partial ^{2}+\xi R}{\mu ^{2}}\right) =4\log \left( \frac{\mu \beta }{2\pi }\right) \zeta _{R}\left( 0\right) +4\zeta ^{\, \prime }\left( 0\right) .
\end{equation}
 The last term is linear in \( \beta  \) and can be discarded, as
we did above (its contribution can readily be absorbed in the regularization
mass \( \mu  \)).

An alternative way (much more elegant mathematically) of treating
the thermal contribution of the zero mode is to proceed along the
lines pioneered in \cite{KE1996}, that is (in a nutshell) to use
the ordinary formulas of zeta regularization till the end, while `lifting'
the zero modes (e.g. preventing them from becoming zero), both the
spatial and temporal one in this case, with the help of two small
parameters (say \( \varepsilon  \) and \( \eta  \)), which are finally
taken to zero after the whole zeta regularization process (which uses
the ordinary expressions in the absence of zero modes) has been carried
through. This method provides, in some situations, a justification
of the usual principal part prescription (see \cite{KE1996}), and
in the present case leads also to a complete cancellation (more than
that, it actually avoids the appearence) of the divergences that we
have just discarded on physical (dimensional) grounds.

Making use of the integral representation for the modified Bessel
function of the third kind \cite{Grad94}\begin{equation}
\int _{0}^{\infty }\, dx\, x^{\nu -1}e^{-\frac{a}{x}-bx}=2\left( \frac{a}{b}\right) ^{\nu /2}K_{\nu }\left( 2\sqrt{ab}\right) ,
\end{equation}
 for the temperature-dependent part describing the non-zero modes,
and adding the zero-mode contribution, we have \begin{equation}
\label{log partition function}
\log \, Z\left( \beta \right) =-\frac{\beta }{2a}\sum _{\ell =0}^{\infty }D_{\ell }M_{\ell }-\frac{1}{2}\log \left( \frac{\mu \beta }{2\pi }\right) -\sum _{\ell =0}^{\infty }D_{\ell }\log \left( 1-e^{-\beta M_{\ell }/a}\right) ,
\end{equation}
 where we have also used the explicit form of \( K_{1/2}\left( z\right)  \)
\cite{Grad94}. Notice that in this approach ---and also in the one
described in \cite{KE1996}--- there is no temperature dependent pole
as in \cite{Dowker84}. The factor \( 2\pi  \) in the denominator
of the log term in this equation is fully arbitrary (it can obviously
be absorbed into the \( \mu  \) parameter). It would lack any sense
to set \( \mu  \) equal to a particular value (say 1, or \( 2\pi  \)),
as \( \mu  \) clearly provides the (regularization) scale.

Equation (\ref{log partition function}) can be applied to several
situations concerning scalar conformally coupled fields on a \( d_{R} \)-dimensional
manifold. The first term of Eq. (\ref{log partition function}) is
associated with the zero temperature vacuum energy, the second one
describes the thermal corrections due to the spatial zero mode (when
present, if not so, this term has just to be deleted), and the third
one is the thermal contribution due to real excitations of the conformally
coupled fields on \( S^{d-1} \). This simple and easy to interpret
formula is the result of the generalised zeta function approach used
here.

\section{The thermodynamics associated with the zero mode}

We consider, to start with, the thermodynamics associated with the
zero mode. First of all, let us remark that the thermal energy associated
with the zero mode does not depend on the scaling mass, in fact\begin{equation}
\label{EZM}
E_{Z.M.}\left( \beta \right) =-\frac{d}{d\beta }\log \, Z_{Z.M.}\left( \beta \right) =\frac{d}{d\beta }\log \, \left( \frac{\mu \beta }{2\pi }\right) =\frac{1}{\beta }.
\end{equation}
 On the contrary, the free energy and the entropy do. The free energy
is\begin{equation}
\label{FZM}
F_{Z.M.}\left( \beta \right) =-\frac{1}{\beta }\log \, Z_{Z.M.}\left( \beta \right) =\frac{1}{\beta }\log \, \left( \frac{\mu \beta }{2\pi }\right) ,
\end{equation}
 and the entropy reads\begin{equation}
\label{SZM}
S_{Z.M.}\left( \beta \right) =-\beta ^{2}\frac{d}{d\beta }\frac{1}{\beta }\log \, Z_{Z.M.}\left( \beta \right) =-\log \left( \frac{\mu \beta }{2\pi }\right) +1.
\end{equation}
 Since the thermal energy and the free energy are related by \begin{equation}
\beta F\left( \beta \right) =\int E\left( \beta \right) +C,
\end{equation}
 we see that, in order to have complete compatibility between (\ref{EZM})
and (\ref{FZM}), we must necessarily choose \( C=\log \left( \mu /2\pi \right)  \).
One can also verify, quite easily, that the fundamental relation between
the three quantities above, \begin{equation}
\label{SEundF}
S=\beta \left( E-F\right) ,
\end{equation}
 is actually satisfied by (\ref{EZM}), (\ref{FZM}) and (\ref{SZM}).
These results are in complete agreement with the ones obtained by
Cardy \cite{Cardy1991}, provided that we choose \( \mu =2\pi  \)
`in inverse length units'. Their physical relevance, however, is dubious
because, as we have already advanced and clearly see now, in order
to maintain internally consistent thermodynamics, the need for a dimensionfull
parameter \( \mu  \) is unavoidable and setting it equal to any particular
value at this stage (that is, setting arbitrarily a mass scale) is
not justified. For this reason, in the following we will leave the
zero mode aside (however, we will discuss the effect of its inclusion
a bit later).

\section{The high-temperature regime}

For conformally coupled scalar fields on \( S^{d-1} \), the relevant
contribution from Eq. (\ref{log partition function}) (skipping the
zero mode) reads \begin{equation}
\label{log thermalpart function}
\log \, Z\left( \beta \right) =-\sum _{\ell =0}^{\infty }\left( \ell +1\right) ^{d-2}\log \left( 1-e^{-\beta \left( \ell +1\right) /a}\right) .
\end{equation}
 And, also from Eq. (\ref{log partition function}), the zero temperature
vacuum energy is\begin{equation}
E_{0}=\frac{1}{2a}\sum _{\ell =0}^{\infty }\left( \ell +1\right) ^{d-2}\left( \ell +1\right) =\frac{1}{2a}\zeta \left( 1-d\right) .
\end{equation}
 Let us consider the one-dimensional compact space for which \( d_{R}=1 \)
and \( D=d=2 \). In this case\begin{equation}
\label{bivac energy}
E_{0}=\frac{1}{2a}\zeta \left( -1\right) =-\frac{1}{24a}.
\end{equation}
 On the other hand, Eq. (\ref{log thermal part function}) reads \begin{equation}
\label{thermalpart}
\log \, Z\left( \beta \right) =-\sum _{\ell =1}^{\infty }\log \left( 1-e^{-\beta \ell /a}\right) ,
\end{equation}
 where we have shifted the quantum number \( \ell  \) by one unit.
Since the thermal energy is given by \begin{equation}
\label{thermal energy}
\widetilde{E}\left( \beta \right) =-\frac{d}{d\beta }\log \, Z\left( \beta \right) ,
\end{equation}
 one can combine (\ref{bivac energy}), (\ref{thermalpart}) and (\ref{thermal energy})
to obtain the total energy, in the form \begin{eqnarray}
E\left( \beta \right)  & = & E_{0}+\widetilde{E}\left( \beta \right) \nonumber \\
 & = & -\frac{1}{24a}+\frac{1}{a}\sum _{\ell =1}^{\infty }\, \frac{\ell }{e^{\beta \ell /a}-1}.\label{Brevik et al 01}
\end{eqnarray}
 One can now make use of an appropriate sum formula, say Euler-MacLaurin,
Abel-Plana, or Poisson, in one of its several versions, to evaluate
the sum on the rhs of (\ref{Brevik et al 01}). As an alternative
to Euler-Maclaurin's formula employed in \cite{Brevik al 02}, let
us consider here the Abel-Plana sum formula\begin{equation}
\label{Abel-Plana}
\sum _{n=1}^{\infty }\, f\left( n\right) =-\frac{1}{2}f\left( 0\right) +\int _{0}^{\infty }\, dx\, f\left( x\right) +i\int _{0}^{\infty }\, dx\, \frac{f\left( ix\right) -f\left( -ix\right) }{e^{2\pi x}-1}.
\end{equation}
 A straightforward application of this expression, with \( f\left( x\right) =x/\left( e^{\beta x/a}-1\right)  \),
leads to \begin{equation}
\label{Brevik et al 02}
E\left( \beta \right) =-\frac{1}{2\beta }+\frac{1}{24a}\left( \frac{2\pi a}{\beta }\right) ^{2},\, \, \, \, \, \, a/\beta \, \gg \, 1,
\end{equation}
 which is the result obtained in \cite{Brevik al 02} and questioned
in \cite{Dowker 2002}. The inclusion of the zero mode here would
have certainly led, unambiguously, to a cancellation of the \( \beta ^{-1} \)
dependence. But this is a very specific feature of the thermal energy,
which is the \textit{only} thermodynamical quantity that does not
depend on the regularization scale of the zero mode.

Notice that (\ref{Brevik et al 01}) is more appropriate for a low-temperature
representation of the energy, and upon the use of the Abel-Plana sum
formula it yields the high-temperature result. As is well known, this
is a property of this kind of sum formulas (starting with the Jacobi
identity), which relate a series expansion valid for say low values
of the relevant parameter with a corresponding series expansion for
large values of the parameter. In spite of those formulas being identities
(both expansions correspond of course to the same function), one should
be careful not to mix terms of one of the expansions with terms of
the other one.

It can be verified precisely that it is the third term in the Abel-Plana
formula the one that exactly cancels the zero temperature contribution.
In fact, the contribution of this term reads\begin{equation}
i\int _{0}^{\infty }\, dx\, \frac{f\left( ix\right) -f\left( -ix\right) }{e^{2\pi x}-1}=\int _{0}^{\infty }\, dx\, \frac{x}{e^{2\pi x}-1}=\frac{1}{24}.
\end{equation}
 Observe also that the result given by (\ref{Brevik et al 02}) is
closed, in the sense that the use of the Abel-Plana sum formula does
not allow for exponential corrections, hence it holds only in the
very high temperature regime.

\section{The low-temperature regime}

We will now construct an alternative representation for the thermal
energy. Contrary to what has been done in the preceding section, through
most of the present one we will be dealing with expressions valid
in the high-temperature regime, and towards the end we will make use
of the Abel-Plana formula in order to obtain the desired low-temperature
expansion.%
\footnote{Both the high and low temperature expansions can be
derived, in a completely equivalent but more direct way, from the
representation of Eq. (\ref{Brevik et al 01}) in terms of the
normalized Eisenstein series \( E_{2}(\tau ) \) \cite{ovm,kobl1}.
}

In order to obtain, to begin with, a high-temperature representation
of the free energy, we make use of the Poisson sum formula, in the
form\begin{equation}
\label{Poisson cosine sum formula}
\sum _{n=1}^{\infty }\, f\left( n\right) =-\frac{f\left( 0\right) }{2}+\int _{0}^{\infty }\, dx\, f\left( x\right) +2\sum _{n=1}^{\infty }\int _{0}^{\infty }\, dx\, f\left( x\right) \, \cos \left( 2\pi nx\right) ,
\end{equation}
 where here \( f\left( x\right) =\log \left( 1-e^{-\beta x/a}\right)  \).
A straightforward application of (\ref{Poisson cosine sum formula})
to (\ref{thermalpart}) yields \begin{equation}
\log \, Z\left( \beta \right) =\frac{f\left( 0\right) }{2}-I\left( 0\right) -2\sum _{n=1}^{\infty }I\left( n\right) ,
\end{equation}
 where\begin{equation}
I\left( n\right) :=\int _{0}^{\infty }\, dx\, f\left( x\right) \, \cos \left( 2\pi nx\right) ,\, \, \, \, \, \, \, n=0,1,2,3,\ldots .
\end{equation}
 The integrals defined by \( I\left( n\right)  \) can be readily
evaluated with the help of formulas to be found in \cite{Grad94},
and the result is\begin{equation}
\log \, Z\left( \beta \right) =\frac{f\left( 0\right) }{2}+\frac{\pi ^{2}a}{6\beta }-\frac{\beta }{24a}+\frac{1}{2}\sum _{n=1}^{\infty }\frac{1}{n}\coth \left( \frac{2\pi ^{2}na}{\beta }\right) .
\end{equation}
 The contribution due to \( f\left( 0\right)  \) can be explicitly
calculated as follows. Expanding the log as usual we can write \begin{equation}
\label{doublepole}
f\left( x\right) =-\frac{1}{2\pi i}\int _{c-i\infty }^{c+i\infty }\, d\alpha \, \Gamma \left( \alpha \right) \zeta \left( \alpha +1\right) \left( \frac{\beta x}{a}\right) ^{-\alpha }
\end{equation}
 where we have made use of the representation\begin{equation}
\label{mt1}
e^{-s}=\frac{1}{2\pi i}\int _{c-i\infty }^{c+i\infty }\, d\alpha \, \Gamma \left( \alpha \right) \, s^{-\alpha },\, \, \, \, \, \, \, \, \, \, \, \, s,c\, \, \mbox {real},\, \, \left| s\right| <1,\, \, c>1.
\end{equation}
 In the limit \( x\rightarrow 0 \) the finite contribution of \( f\left( x\right)  \)
is determined by the double pole of the integrand in Eq. (\ref{doublepole})
at \( \alpha =0 \), \begin{equation}
\lim _{x\rightarrow 0}\, f\left( x\right) =f\left( 0\right) =\log \left( \frac{\beta }{2\pi a}\right) .
\end{equation}
 On the other hand, we can also make use of the expansion\begin{equation}
\coth \left( x\right) =1+2\sum _{p=1}^{\infty }e^{-2px},\, \, \, \, \, \, \, \, x>0.
\end{equation}
 As a consequence, after discarding an unphysical (formally divergent)
constant we end up with \begin{equation}
\log \, Z\left( \beta \right) =\frac{\pi ^{2}a}{6\beta }+\frac{1}{2}\log \left( \frac{\beta }{2\pi a}\right) -\frac{\beta }{24a}+\sum _{p,\, n=1}^{\infty }\frac{1}{n}\, e^{-4\pi ^{2}\, n\, p\, a/\beta }.
\end{equation}
 It follows that the thermal part of the free energy in this representation
is given by\begin{eqnarray}
\widetilde{F}\left( \beta \right)  & = & -\frac{1}{\beta }\log \, Z\left( \beta \right) \nonumber \\
 & = & -\frac{\pi ^{2}a}{6\beta ^{2}}-\frac{1}{2\beta }\log \left( \frac{\beta }{2\pi a}\right) +\frac{1}{24a}-\frac{1}{\beta }\sum ^{\infty }_{p,\, n=1}\frac{1}{n}\, e^{-4\pi ^{2}\, n\, p\, a/\beta }.\label{nkl1}
\end{eqnarray}
 As in the case of Sect. 2, an alternative calculation can be performed
using standard zeta function methods \textit{ab initio}, namely, the
transform (\ref{mt1}) already in the expansion of the starting expression
(\ref{log thermal part function}). The final result obtained is exactly
the same, but the calculation is very much shortened and (as before)
it avoids at every step the formation of divergences.%
\footnote{We owe K. Kirsten this observation and the correction of an error
in our previous result.
}

Notice that our result, Eq. (\ref{nkl1}), has a log term and therefore
does not fulfill the general relations given in \cite{Kutasov and Larsen 2000},
namely: that the high temperature expansion of the free energy must
have the general form \begin{equation}
-\widetilde{F}\left( \beta \right) a=a_{d}\left( \frac{2\pi a}{\beta }\right) ^{d}+a_{d-1}a_{d}\left( \frac{2\pi a}{\beta }\right) ^{d-2}+\cdots +a_{0}\left( \frac{2\pi a}{\beta }\right) ^{0}+O\left( e^{-4\pi ^{2}a/\beta }\right) ,
\end{equation}
 the coefficients of this expansion satisfying the relation \begin{equation}
\sum _{k=0}^{d/2}\left( -1\right) ^{k}\left( 2k-1\right) a_{2k}=0.
\end{equation}
 If not for the log term, these relations would be fulfiled. Observe,
however, that there would not be an immediate cancellation of the
log term if we had included the zero mode contribution. In fact, the
matching of the zero mode term, \( \log \left( \frac{\mu \beta }{2\pi }\right)  \),
with the one, with opposite sign, coming from the high-\( T \) limit,
-\( \log \left( \frac{\beta }{2\pi a}\right)  \), is not a triviality.
On the contrary, it has a very deep meaning: it takes place if and
only if we \textit{prescribe} that the regulating mass-dimensional
parameter be set exactly equal to the inverse of the compactification
length, \( \mu =1/a \). This may certainly be viewed as a most natural
choice and a physically meaningful one ---being \( 1/a \) the only
caracteristic dimension in our model--- but even then, this is no
substitute for the pertinent renormalization procedure.

The total free energy is\begin{equation}
F\left( \beta \right) =E_{0}+\widetilde{F}\left( \beta \right) =-\frac{\pi ^{2}a}{6\beta ^{2}}-\frac{1}{2\beta }\log \left( \frac{\beta }{2\pi a}\right) -\frac{1}{\beta }\sum ^{\infty }_{p,\, n=1}\frac{1}{n}\, e^{-4\pi ^{2}\, n\, p\, a/\beta }.
\end{equation}
 The total energy reads\begin{equation}
E\left( \beta \right) =\frac{d}{d\beta }\left( \beta F\left( \beta \right) \right) =\frac{\pi ^{2}a}{6\beta ^{2}}-\frac{1}{2\beta }-\frac{4\pi ^{2}a}{\beta ^{2}}\sum ^{\infty }_{p,\, n=1}\, p\, \, e^{-4\pi ^{2}n\, p\, a/\beta },
\end{equation}
 and can be summed over \( n \) to yield the high temperature representation\begin{equation}
E\left( \beta \right) =\frac{\pi ^{2}a}{6\beta ^{2}}-\frac{1}{2\beta }-\frac{4\pi ^{2}a}{\beta ^{2}}\sum ^{\infty }_{p=1}\, \frac{p}{e^{-4\pi ^{2}pa/\beta }-1}.
\end{equation}
 If we now, as before, make use of the Abel-Plana sum formula (\ref{Abel-Plana}),
we obtain\begin{equation}
E\left( \beta \right) =-\frac{1}{24a},\, \, \, \, \, \, \, \, \, a/\beta \, \ll \, 1\, .
\end{equation}
 Notice that, as anounced above, we have here started from a high-temperature
representation for the energy and now, upon the use of the Abel-Plana
sum formula, we obtain the energy in the zero-temperature limit, a
procedure that can be considered the reciprocal of the one employed
in Ref. \cite{Brevik al 02}. Notice also that in the Abel-Plana formula,
the first term leads to the zero temperature term, the second cancels
out the linear contribution and the third one cancels out the Stefan-Boltzmann
term.

\section{Final remarks}

In this brief analysis, we have focused on the possible contribution
of the spatial zero mode to the thermal energy of a scalar field in
a \( S^{1}\times S^{1} \) geometry. We have shown that, though the
thermal energy does not depend on the scaling mass \( \mu  \), the
partition function, the free energy and the entropy do depend on the
choice of value for this parameter. The important obstruction to be
circumvented is to satisfy the third law of thermodynamics for, as
it stands, the entropy (\ref{SZM}) associated with the spatial zero
mode diverges in the zero temperature limit, a fact also noticed in
\cite{Brevik al 02}. It was argued in \cite{Dowker84} that the scaling
mass can be redefined. Unfortunately, any tentative of redefining
the scaling mass, that is, of eliminating the scaling mass dependence
from (\ref{SZM}), will lead to clear contradictions among the thermodynamical
relations linking the basic quantities (\ref{EZM}), (\ref{FZM})
and (\ref{SZM}). In fact, the scaling mass parameter cannot be made
disappear. At most, it can be traded for a characteristic mass or
inverse length of the system considered (here \( 1/a \), the inverse
compactification radius). A subsequent renormalization procedure seems
inescapable, which makes contact in the end with physical reality.

Let us point out that, although we have here considered a very simple
model, in order to keep all terms under easy control and render the
argument clear, the results obtained are valid under much more general
circumstances, affecting very general physical systems that develop
zero modes. The treatment has been that of (a natural extension of)
the zeta function regularization prescription.

To sum up, only if one restricts the analysis to the thermal energy
---and nothing else--- can one avoid the situation here presented,
since this is the only thermodynamical quantity that does not
depend on the scaling mass. It would seem, thus, that coherence
forces one to set the zero mode and its thermodynamics aside.
Notice also that the thermal energy, the entropy and the free
energy of the zero mode do not explicitly depend on the radius (or
characteristic compactification length) of the manifold. This
means that (at least at the one-loop level) their contribution to
any isothermal process ---e.g., the only ones permitted if the
system is in thermal equilibrium with a heath bath--- will not be
physically observable. But this consideration would be completely
reversed if the necessity for setting in fact \( \mu =1/a \) would
prevail in the end (as one of the factions in the already
mentioned controversy tend to support), and this would open the
door to an experimental manifestation of the zero mode and to the
possibility of answering in the affirmative the important question
of the controverted physical relevance of these zero modes. In
that sense, a fine experimental validation of Eq. (\ref{Brevik et
al 02}) would be crucial (presence or absence of the linear
dependence). For what we have seen, this would imply, necessarily,
the breaking of the thermodynamical relations and the necessity of
reformulating the third principle in the quantum field theory
domain.

\section*{Acknowledgments}

The authors are indebted with K. Kirsten for several valuable observations.
A.C.T. wishes to acknowledge the hospitality of the Institut d'Estudis
Espacials de Catalunya (IEEC/CSIC) and the Universitat de Barcelona,
Departament d'Estructura i Constituents de la Mat\`{e}ria and the
financial support of CAPES, the Brazilian agency for faculty improvement,
Grant BExt 0168-01/2. The investigation of E.E. has been supported
by DGI/SGPI (Spain), project BFM2000-0810, and by CIRIT (Generalitat
de Catalunya), contract 1999SGR-00257.

\end{document}